\documentclass[preprint,12pt,a4paper]{elsarticle}

\usepackage{multirow}

\usepackage{epsfig}

\usepackage{amssymb}

\usepackage{float}

\usepackage{booktabs} 

\usepackage{comment}
\usepackage{ulem}

\newcommand{\tcs}{$\sigma_{tot}$}
\newcommand{\ro}{$\rho$}

\biboptions{compress}

\begin{document}

\begin{frontmatter}

\title{Bounds on the rise of total cross section from LHC7 and LHC8 data}



\author[label1]{D. A. Fagundes}
\author[label2]{M. J. Menon}
\author[label2]{P. V. R. G. Silva}

\address[label1]{Departamento de Ci\^encias Exatas e Educa\c c\~ao, Universidade Federal de Santa Catarina - 
Campus Blumenau, 89065-300 Blumenau, SC, Brazil}
\address[label2]{Instituto de F\'{\i}sica Gleb Wataghin,
Universidade Estadual de Campinas - UNICAMP\\
13083-859 Campinas, SP, Brazil}

\begin{abstract} 
Recent measurements of the proton-proton total cross section $\sigma_{tot}$ at 7 and 8 TeV
by the TOTEM and ATLAS Collaborations are characterized by some discrepant values:
the TOTEM data suggest a rise of the cross section with the energy faster than
the ATLAS data. Attempting to quantify these different behaviors, we develop 
new analytical fits to $\sigma_{tot}$ and $\rho$ data from $pp$ and $\bar{p}p$
scattering in the energy region 5 GeV - 8 TeV. 
The dataset comprises all the accelerator data below
7 TeV and we consider three ensembles by adding: either only the TOTEM data
(T), or only the ATLAS data (A), or both sets (T+A).
For the purposes, we use our previous RRPL$\gamma$ parametrization for $\sigma_{tot}(s)$,
consisting of two Reggeons (RR), one critical Pomeron (P) and a leading log-raised-to-gamma
(L$\gamma$) contribution (with $\gamma$ as a free fit parameter), \textit{analytically} connected to
$\rho(s)$ through singly-subtracted derivative dispersion relations and energy scale
fixed at the physical threshold.  The data reductions with ensembles
T and A present good agreement with the experimental data analyzed and cannot be distinguished
on statistical grounds. The quality of the fit is not as good with ensemble T+A.
The fit results provide
$\gamma \sim 2.3 \pm 0.1$ (T), $2.0 \pm 0.1$ (A), $2.2 \pm 0.2$ (T+A),
with $\chi^2/\mathrm{DOF} \sim 1.07$ (T), $1.09$ (A), $1.14$ (T+A),
suggesting extrema bounds for $\gamma$ given by 1.9 and 2.4. Fits with $\gamma = 2$ (fixed) are also developed
and discussed.
\end{abstract}


\end{frontmatter}

\vspace{0.5cm}

\centerline{\textit{Published in Nuclear Physics A (2017)}}

\vspace{0.5cm}


Contents

\vspace{0.1cm}

1. Introduction


2. Analytical Parametrization


3. Ensembles, Fits Procedures and Fit Results

\ \ \ \ 3.1 Experimental Data and Ensembles

\ \ \ \ 3.2 Fit Procedures 

\ \ \ \ 3.3 Fit Results


4. Summary, Conclusions and Final Remarks


Appendix A. Comments on the Dataset and Fit Procedures


\section{Introduction}
\label{s1}

The elastic scattering of hadrons at high energies is characterized by 
its amplitude $A$ as a function of the c.m. energy and transferred 
momentum squared in the collision process, $s$ and $t$. These soft scattering states
constitute one of the great challenges for the QCD, because, 
as a long distance phenomena, its investigation
demands a nonperturbative approach and presently, a formal approach from first principles, 
able to provide $A = A(s,t)$, is still absent.

In the forward scattering, the elastic amplitude is connected with two important physical
observables: the total cross section (optical theorem),
\begin{eqnarray}
\sigma_{tot}(s) = \frac{\mathrm{Im}\,A(s,t=0)}{s}
\label{e1}
\end{eqnarray}
and the $\rho$ parameter (related to the phase of the amplitude),
\begin{eqnarray}
\rho(s) = \frac{\mathrm{Re}\,A(s,t=0)}{\mathrm{Im}\,A(s,t=0)}.
\label{e2}
\end{eqnarray}

A fundamental formal result on the rise of $\sigma_{tot}(s)$, at asymptotic energies
($s \rightarrow \infty$), is the upper bound by
Froissart, Lukaszuk, Martin \cite{froissart,martin1,martin2,lukamartin}
\begin{eqnarray}
\sigma_{tot}(s) < c \ln^2(s/s_0),
\label{e3}
\end{eqnarray}
where $s_0$ is an unspecified energy scale and the pre-factor on the right-hand side is bounded by
$\pi/m_{\pi}^2$, where $m_{\pi}$ is the pion mass \cite{lukamartin}.

Beyond specific phenomenological models, an important approach in the investigation
of $\sigma_{tot}(s)$ and $\rho(s)$ is the forward amplitude analysis, characterized
by Regge-Gribov analytical parametrization for $\sigma_{tot}(s)$, connected with 
$\rho(s)$ by means of dispersion relations or asymptotic uniqueness and analytical or numerical
methods (for example, \cite{amaldi,ua42,ckk,compete1,compete2,lm,ii,bh, pdg14,pdg16}).

A typical behavior emerging from several analyses favors a leading $\ln^2 s$ dependence
for $\sigma_{tot}(s)$ at the highest energies \cite{compete1,compete2,ii,bh,pdg14,pdg16}.
In a nonperturbative QCD approach this behavior has been also recently predicted under
specific conditions \cite{gm1,gm2}.

Some authors have considered a leading $\ln^{\gamma} s$ dependence, with $\gamma$
as a free fit parameter. Different approaches using different datasets indicated
$\gamma$ = 2.10 $\pm$ 0.10 \cite{amaldi} (1977), 2.24$^{+0.35}_{-0.31}$ \cite{ua42} (1993)
and in our more recent analyses, $\gamma$ = 2.2 - 2.4 \cite{fms1,fms2,ms1,ms2},
with the last updated result $\gamma$ = 2.23 $\pm$ 0.11 \cite{ms2} (2013).
In the theoretical context, the possibility of a rise of $\sigma_{tot}(s)$ faster
than $\ln^2 s$, without violating unitarity, is discussed in \cite{azimov}.
It should be also noted that Eqs. (3-4) are intended for $s \rightarrow \infty$ and presently,
experimental data from accelerators have been obtained up to $\sqrt{s} =$ 8 TeV. 
To a certain extent contrasting with the above $\gamma$ values, the COMPAS group (PDG) has quoted the value
1.98 $\pm$ 0.01 \cite{pdg14,pdg16}.

In the experimental context, the recent data from LHC7 and LHC8 are expected to provide
fundamental information for amplitude analyses. 
These new experimental information comprise 5 measurements of $\sigma_{tot}$ 
at 7 TeV (4 data obtained by the TOTEM Collaboration \cite{totem1,totem2,totem3}
and 1 datum by the ATLAS Collaboration \cite{atlas7}), 6 measurements at 8 TeV
(5 data by TOTEM \cite{totem4,totem5,totem6} and 1 datum by ATLAS \cite{atlas8}) and one measurement of
$\rho$ at 8 TeV by the TOTEM Collaboration \cite{totem6}.
All these total cross section data are displayed in Table \ref{t1},
where we have expressed as systematic also the uncertainties associated with extrapolations and error propagated
from uncertainties in the fit parameters.

\begin{table}[ht]
\centering
\caption{Measurements of the $pp$ total cross section from LHC7 and LHC8:
central value ($\sigma_{tot}$), statistical uncertainties
($\Delta\sigma_{tot}^{stat.}$), systematic uncertainties
($\Delta\sigma_{tot}^{syst.}$), total uncertainty from quadrature
($\Delta \sigma_{tot}$).}   
\vspace{0.1cm}
\begin{tabular}{c c c c c c}
\hline 
$\sqrt{s}$ &\tcs &$\Delta\sigma_{tot}^{stat.}$&
$\Delta\sigma_{tot}^{syst.}$&$\Delta \sigma_{tot}$& Collaboration\\
(TeV)& (mb) & (mb) & (mb) & (mb) & [reference] \\
\hline
7.0 & 98.3  & 0.2   & 2.8             & 2.8  & TOTEM \cite{totem1}\\
    & 98.6  &  -    & 2.2             & 2.2  & TOTEM \cite{totem2}\\
    & 98.0  &  -    & 2.5             & 2.5  & TOTEM \cite{totem3}\\
    & 99.1  &  -    &  4.3            & 4.3  & TOTEM \cite{totem3}\\
    & 95.35 & 0.38  &1.304            & 1.36 & ATLAS \cite{atlas7}\\
8.0 & 101.7 & -     & 2.9             & 2.9  & TOTEM \cite{totem4}\\
    & 101.5 & -     & 2.1             & 2.1  & TOTEM \cite{totem5}\\
    & 101.9 & -     & 2.1             & 2.1  & TOTEM \cite{totem5}\\
    & 102.9 & -     & 2.3             & 2.3  & TOTEM \cite{totem6}\\
    & 103.0 & -     & 2.3             & 2.3  & TOTEM \cite{totem6}\\
    & 96.07 & 0.18  & 0.85 $\pm$ 0.31 & 0.92 & ATLAS \cite{atlas8}\\
\hline 
\end{tabular}
\label{t1}
\end{table}

However, TOTEM and ATLAS data present discrepant values. In special, at 8 TeV
the ATLAS datum \cite{atlas8} and the latest measurement by the TOTEM Collaboration 
\cite{totem6} differ by
\begin{eqnarray} 
\frac{\sigma_{tot}^{\mathrm{TOTEM}} - \sigma_{tot}^{\mathrm{ATLAS}}}{\Delta \sigma_{tot}^{\mathrm{TOTEM}}} = 
\frac{103 - 96.07}{2.3} = 3.0.
\nonumber
\end{eqnarray} 
This ratio results 7.5 if the ATLAS uncertainty is considered.

In this situation, it may be important and instructive to attempt 
to quantify these discrepancies through an amplitude analysis, able to test 
statistical consistency with experimental
data at lower energies, as well as estimating uncertainty regions for extrapolations
to higher (and asymptotic) energies.
In this sense, the log-raised-to-gamma leading term provides a useful tool
through the $\gamma$ values (and the pre-factor values) that can be determined from
different fits to different datasets\footnote{Another useful parameter is the soft
Pomeron intercept, related to the simple pole leading contribution, $s^{\epsilon}$,
as done in \cite{lm} in case of $\bar{p}p$ scattering at 1.8 TeV and cosmic-ray
estimations for $\sigma_{tot}$.}. Moreover, the extracted $\gamma$ values
provide also information on the proximity and/or consistency with the particular case $\gamma = 2$ 
(Froissart, Lukaszuk, Martin bound).

To this end, we develop here new fits with our RRPL$\gamma$ parametrization for $\sigma_{tot}(s)$,
analytically connected to
$\rho(s)$ through singly-subtracted derivative dispersion relations \cite{fms1,fms2,ms1,ms2}.
We consider either $\gamma$ as a free fit parameter or $\gamma$ = 2 (fixed).
The dataset comprises all the accelerator data from $pp$ and $\bar{p}p$ elastic scattering
above 5 GeV and up to 8 TeV. For $\sigma_{tot}$, including the data below 7 TeV, 
we consider three ensembles by adding: either only the TOTEM data
(ensemble T), or only the ATLAS data (ensemble A), or both sets (ensemble T+A).

Our main conclusions are the following: (1) the data reductions with ensembles
T or A present good agreement with the experimental data analyzed and cannot be distinguished
on statistical grounds; (2) the quality of the fit is not as good in case of ensemble T+A
and the curves lies between the TOTEM and ATLAS data;
(3) the fits with $\gamma$ free or fixed to 2 are also indistinguishable on statistical grounds;
(4) we infer upper and lower bounds for $\gamma$ given by 2.4 and 1.9. 

The article is organized as follows. In Section \ref{s2} present our analytical parametrization.
In Section \ref{s3} we refer to the experimental data
analyzed (defining the three ensembles) and present the fits procedures and results.
Our conclusions and final remarks are the contents of Section \ref{s4}.

\section{Analytical Parametrization}
\label{s2}

This analytical model has been introduced in 
\cite{fms1} and further developed, extended and discussed in \cite{fms2,ms1,ms2}.
It consists of two assumptions related to $\sigma_{tot}(s)$ and $\rho(s)$.

The total cross section is given by the
parametrization introduced by Amaldi et al. in the 1970s \cite{amaldi}:
\begin{equation}
\sigma_{tot}(s) = a_1\, \left[\frac{s}{s_0}\right]^{-b_1} + 
\tau\, a_2\, \left[\frac{s}{s_0}\right]^{-b_2}
+  \alpha + \beta \ln^{\gamma}\left(\frac{s}{s_0}\right),
\end{equation}
where $\tau$ = -1 (+1) for $pp$ ($\bar{p}p$) scattering and
$a_1$, $b_1$, $a_2$, $b_2$, $\alpha$, $\beta$, $\gamma$ are real free fit parameters.
Here, the energy scale is fixed at the physical threshold (see \cite{ms2}
for discussions on this choice):
\begin{eqnarray}
s_0 = 4m_p^2 \sim \mathrm{3.521\ GeV}^2.
\nonumber
\end{eqnarray}

The $\rho(s)$ dependence is analytically determined through singly subtracted
derivative dispersion relations, using the Kang and Nicolescu representation \cite{kn}
and reads:
\begin{eqnarray}
\rho(s) &=& \frac{1}{\sigma_{\mathrm{tot}}(s)}
\left\{ \frac{K_{eff}}{s} + T_{R}(s) + T_{P}(s) \right\},
\label{e9}
\end{eqnarray}
where $K_{eff}$ is an \textit{effective subtraction constant} (discussed below)
and the terms ($T$) associated with Reggeon ($R$) and Pomeron ($P$) contributions are
given by
\begin{eqnarray}
T_{R}(s) =
- a_1\,\tan \left( \frac{\pi\, b_1}{2}\right) \left[\frac{s}{s_0}\right]^{-b_1} +
\tau \, a_2\, \cot \left(\frac{\pi\, b_2}{2}\right) \left[\frac{s}{s_0}\right]^{-b_2}, 
\label{e10}
\end{eqnarray}
\begin{eqnarray}
T_{P}(s) =
\mathcal{A}\,\ln^{\gamma - 1} \left(\frac{s}{s_0}\right) +
\mathcal{B}\,\ln^{\gamma - 3} \left(\frac{s}{s_0}\right) +
\mathcal{C}\,\ln^{\gamma - 5} \left(\frac{s}{s_0}\right),
\label{e13}
\end{eqnarray}
where
\begin{eqnarray} 
\mathcal{A} = \frac{\pi}{2} \, \beta\, \gamma,  
\quad 
\mathcal{B} = \frac{1}{3} \left[\frac{\pi}{2}\right]^3 \, \beta\, \gamma\, [\gamma - 1][ \gamma - 2], 
 \nonumber \\
\mathcal{C} = \frac{2}{15} \left[\frac{\pi}{2}\right]^5 \, \beta\, \gamma\, [\gamma - 1][ \gamma - 2]
[\gamma - 3][ \gamma - 4].
\label{e14}
\end{eqnarray} 

We refer to the subtraction constant as \textit{effective} because, as a free fit parameter,
it takes account of the constant (not zero) lower limit in the Integral Dispersion Relation,
which is replaced by the derivative form. In a \textit{high-energy approximation} this lower
limit, given by the energy threshold $s_0 = 4m_p^2$, is assumed as zero. However,
once taken into account, it results in a series expansion whose first order term (1/s) can
be absorbed in the $K/s$ term in the dispersion relation, defining our 
\textit{effective subtraction constant}. As a consequence, its introduction means that,
in first order, the high-energy approximation is not included, which gives support
for fits at intermediated and low energies (above the energy cutoff, $\sqrt{s_{min}}$ = 5 GeV).
These aspects related to the role and practical
applicability of the subtraction constant as a free fit parameter are discussed in
\cite{fms2}, Section 2.3.2, \cite{ms2}, Section A.2 and \cite{am}, Section 4.4.

In the particular case of $\gamma = 2$, from Eq. (8), 
$\mathcal{A} = \pi \beta$, $\mathcal{B} = \mathcal{C} = 0$ 
and from Eqs. (4-7), the expressions for \tcs($s$) and \ro($s$) have the \textit{same analytical
structure} as those selected by the COMPETE Collaboration \cite{compete1,compete2} and used
in the successive editions by the PDG \cite{pdg14,pdg16}, except for the presence here of the
effective subtraction constant and the fixed energy scale (not associated
with a free parameter). Also, our pre-factor $\beta$ is not related to
the energy scale.

Inspired by the COMPETE notation \cite{compete1} we can express these two cases
as RRPL$\gamma$ and RRPL2 models. Given the same sub-leading terms, for short,
we shall refer to them only as L$\gamma$ and L2 models.

\section{Ensembles, Fits Procedures and Fit Results}
\label{s3}

\subsection{Experimental Data and Ensembles}

Our dataset on $\sigma_{tot}$ and $\rho$ comprises all the accelerator data from $pp$ and $\bar{p}p$
elastic scattering above 5 GeV \cite{pdgdata} (same cutoff used in the COMPETE and PDG analyses),
including all published results from LHC7 and LHC8 by the TOTEM and ATLAS Collaborations (Table 1).
The recent measurement of \ro\ at 8 TeV by the TOTEM
Collaboration \cite{totem6}
is also included in the dataset. Although not taking part in the data reductions,
we will display in the figures, as illustration, some estimations of \tcs\ from cosmic-ray experiments:
ARGO-YBJ results at $~\,$ 100 - 400 GeV \cite{argo}, Auger result at 57 TeV \cite{auger} 
and Telescope Array (TA) result at 95 TeV \cite{ta}.

We consider three ensembles for $\sigma_{tot}$, all of them including  the data below 7 TeV and
distinguished by adding: only the TOTEM data (ensemble T), or only the ATLAS data (ensemble A),
or both datasets (ensemble T+A).

\subsection{Fit Procedures}

The nonlinearity of the fit demands the choice of initial values for the free parameters
and that is an important point.
As in our previous analyses, we develop data reductions with both the L2 and L$\gamma$ models defined 
in Section 2: first with L2 and then with L$\gamma$ (by letting free the parameter $\gamma$)
and using as initial values for this last fit the results of the former.
For L2 we use as initial values for the free parameters 
the last result by the COMPAS group from PDG 2016 \cite{pdg16}:
\begin{eqnarray}
a_1 &=& \mathrm{13.07\ mb}, \quad b_1 = \mathrm{0.4473}, \quad 
a_2 = \mathrm{7.394\ mb}, \quad b_2 = \mathrm{0.5486},  \nonumber \\
\alpha &=& \mathrm{34.41\ mb}, \quad \beta = \mathrm{0.272\ mb}.
\nonumber
\end{eqnarray}
Since the subtraction constant is absent in the COMPAS parametrization
we consider the initial value 0 for $K_{eff}$.

The above procedure for feedbacks (PDG 2016 $\rightarrow$ L2 $\rightarrow$ L$\gamma$) is applied to each one of the three
ensembles.

As in our previous analysis, 
the data reductions have been performed with the objects of the class TMinuit of ROOT Framework 
\cite{root}. We have employed the default MINUIT error analysis \cite{minuit}
with the \textit{selective criteria} explained in \cite{ms1} (section 2.2.4).
Variances and covariances associated with each free parameter
are used in the analytic evaluation of the uncertainty regions 
associated with the fitted and predicted
quantities.
As tests of goodness-of-fit we shall consider the chi-square per degree of freedom
($\chi^2/\nu$) and
the corresponding integrated probability, $P(\chi^2)$ \cite{bev}.

\subsection{Fit Results}

\begin{figure}[H]
\centering
\epsfig{file=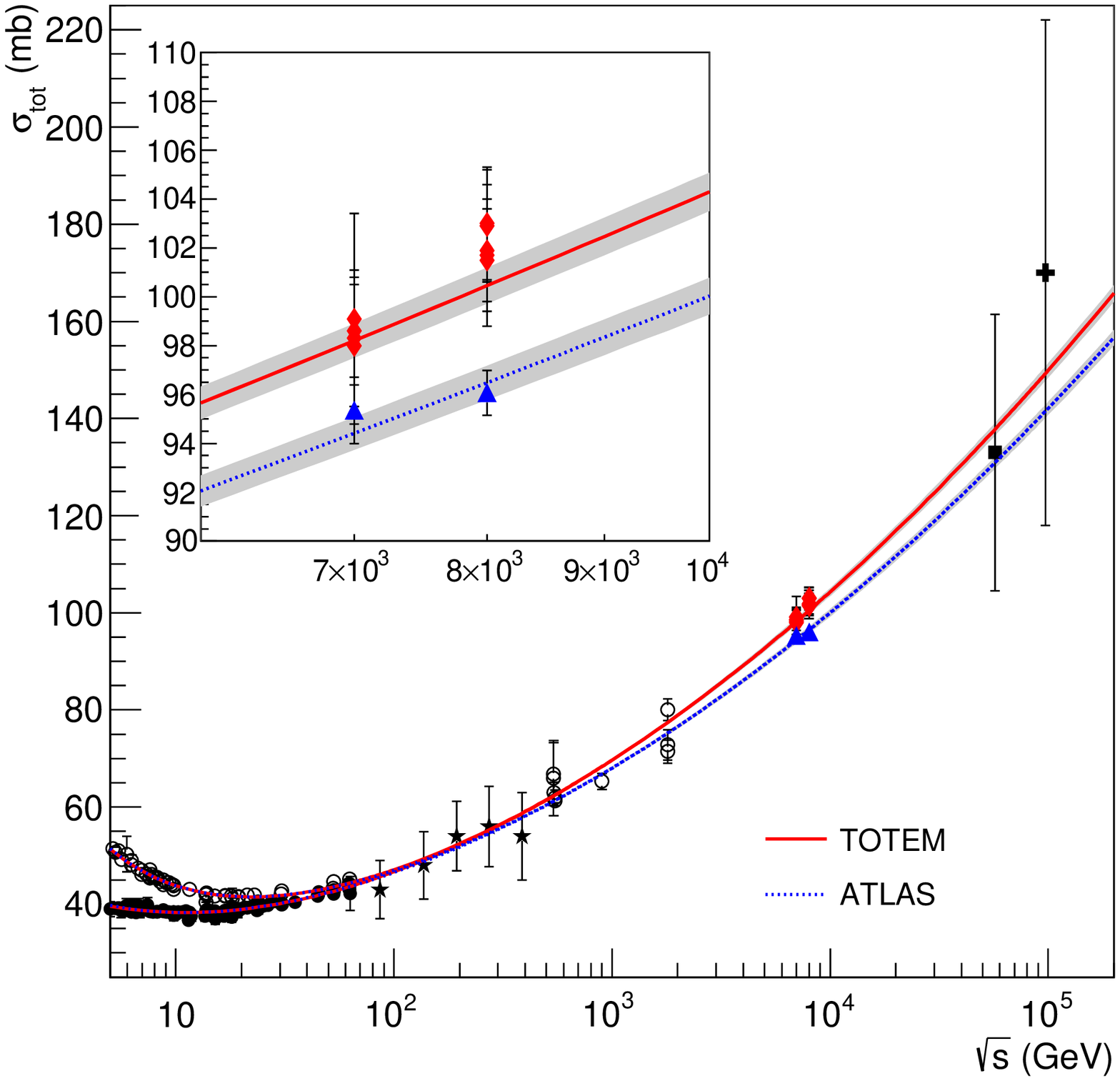,width=8cm,height=8cm}
\epsfig{file=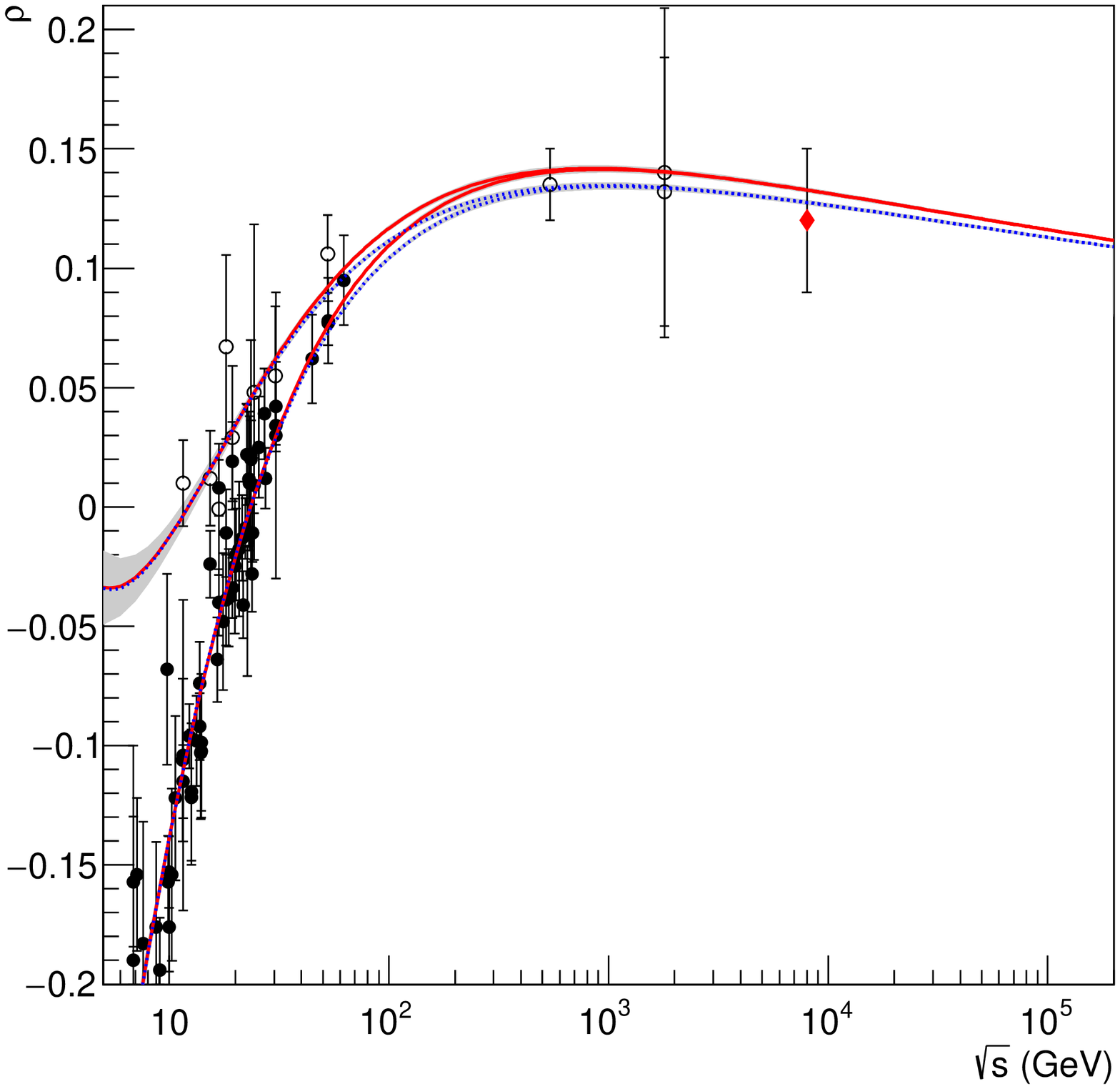,width=8cm,height=8cm}
\epsfig{file=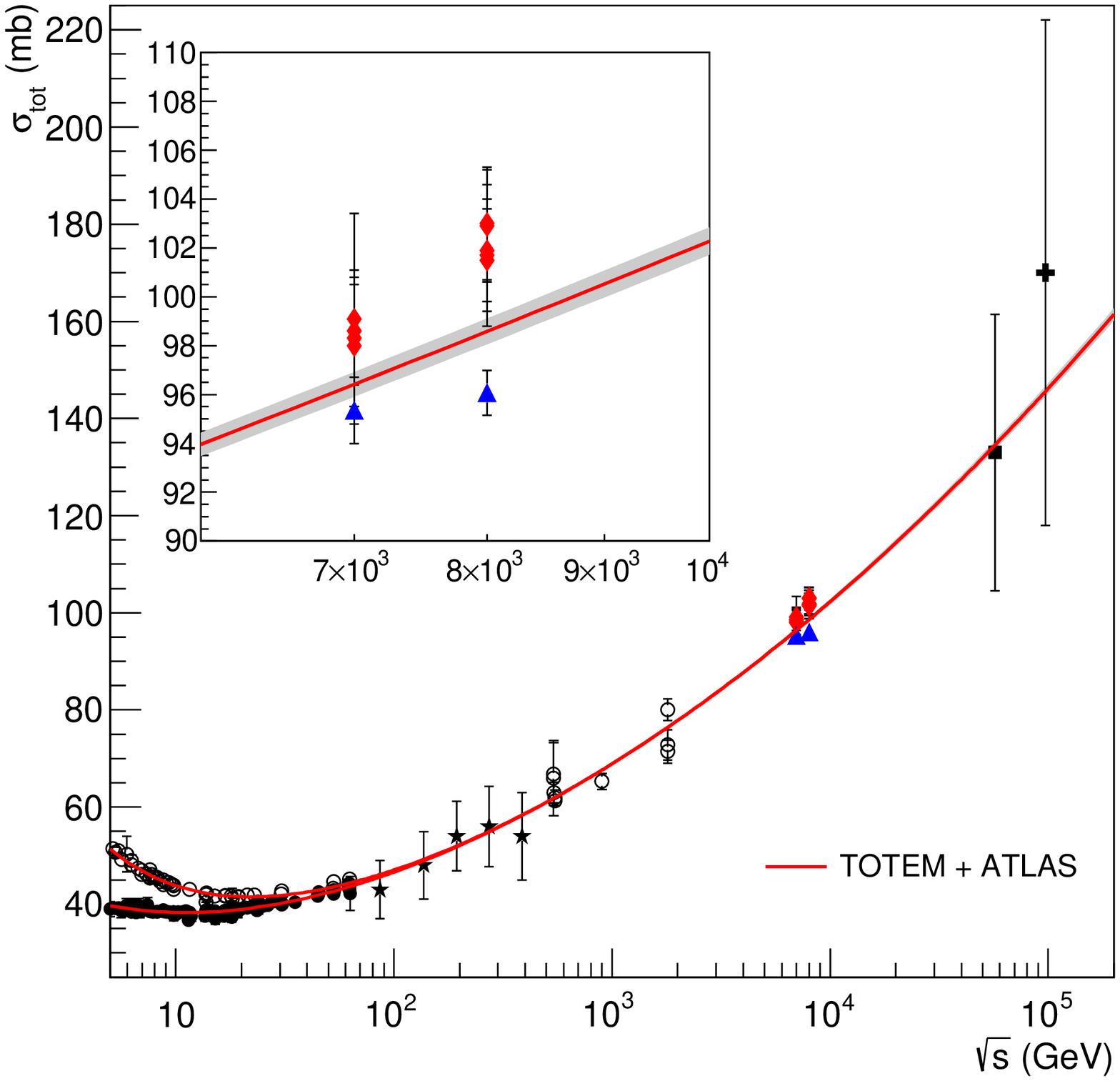,width=8cm,height=8cm}
\epsfig{file=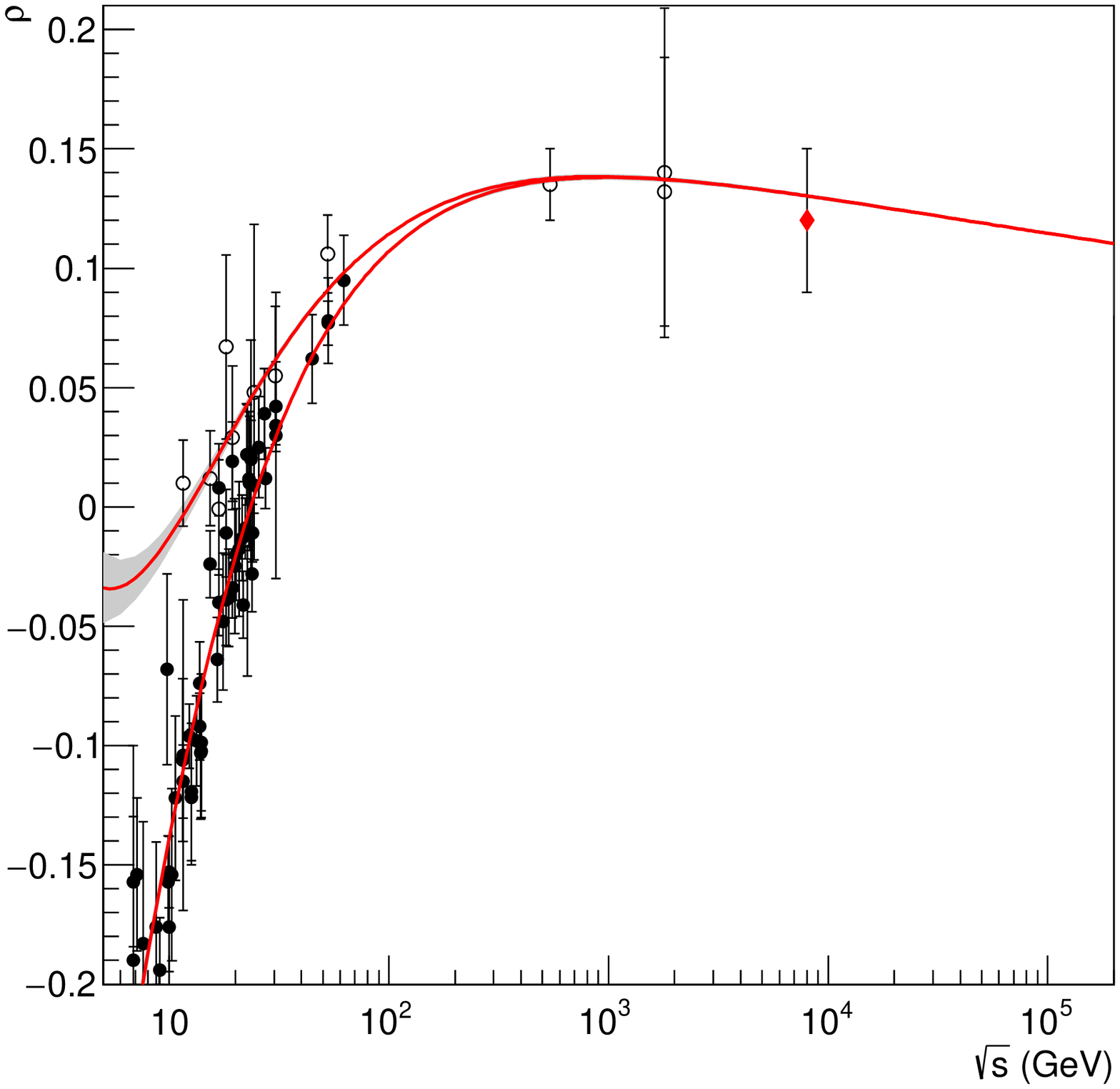,width=8cm,height=8cm}
\caption{Data reductions to $\sigma_{tot}$ (left) and $\rho$ (right) from $pp$ (black circles)
and $\bar{p}p$ (white circles) scattering. Fit results with the \textbf{L2 model}
to ensembles T and A (above) and T+A (below).}
\label{f1}
\end{figure}

\begin{figure}[H]
\centering
\epsfig{file=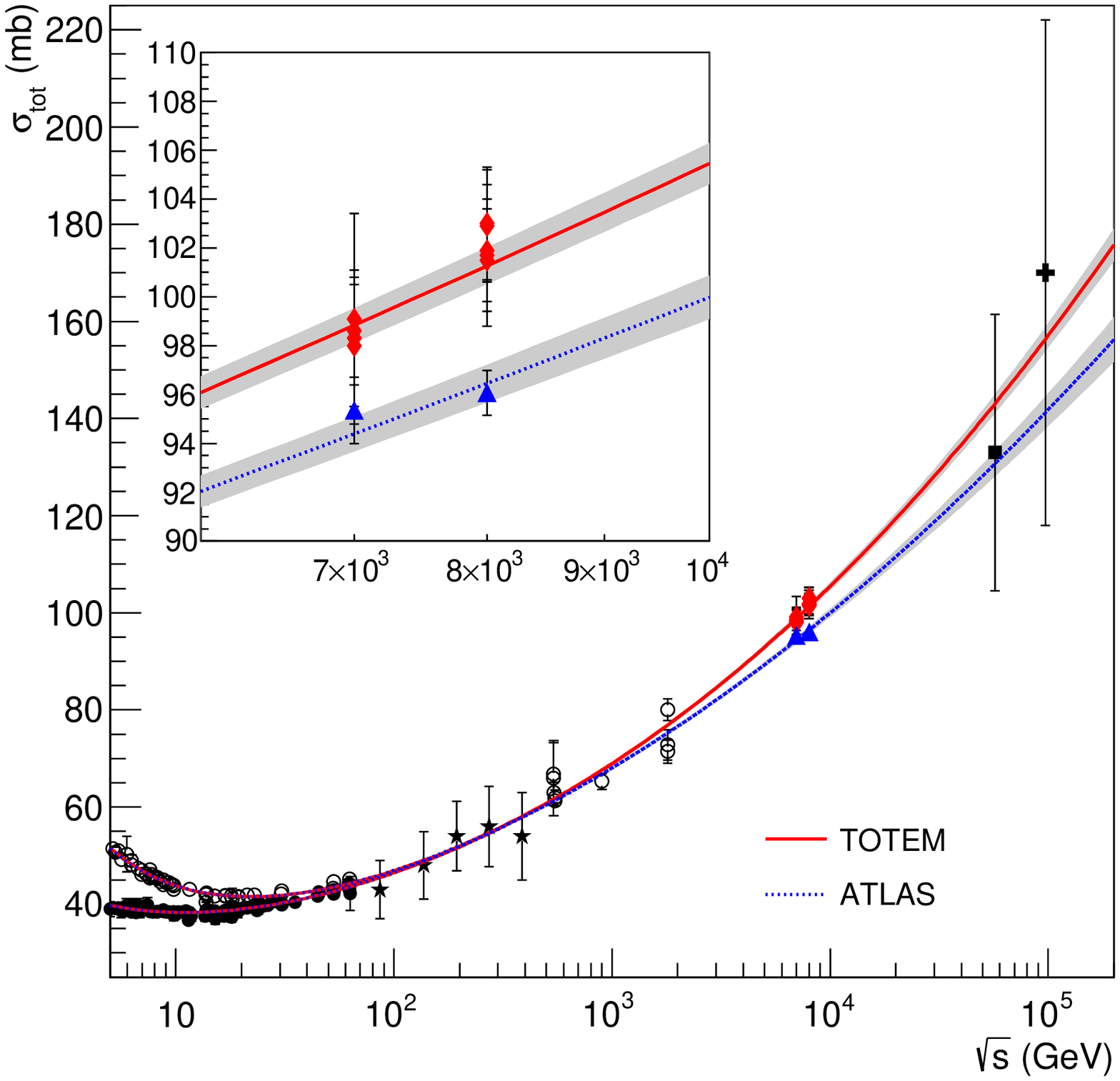,width=8cm,height=8cm}
\epsfig{file=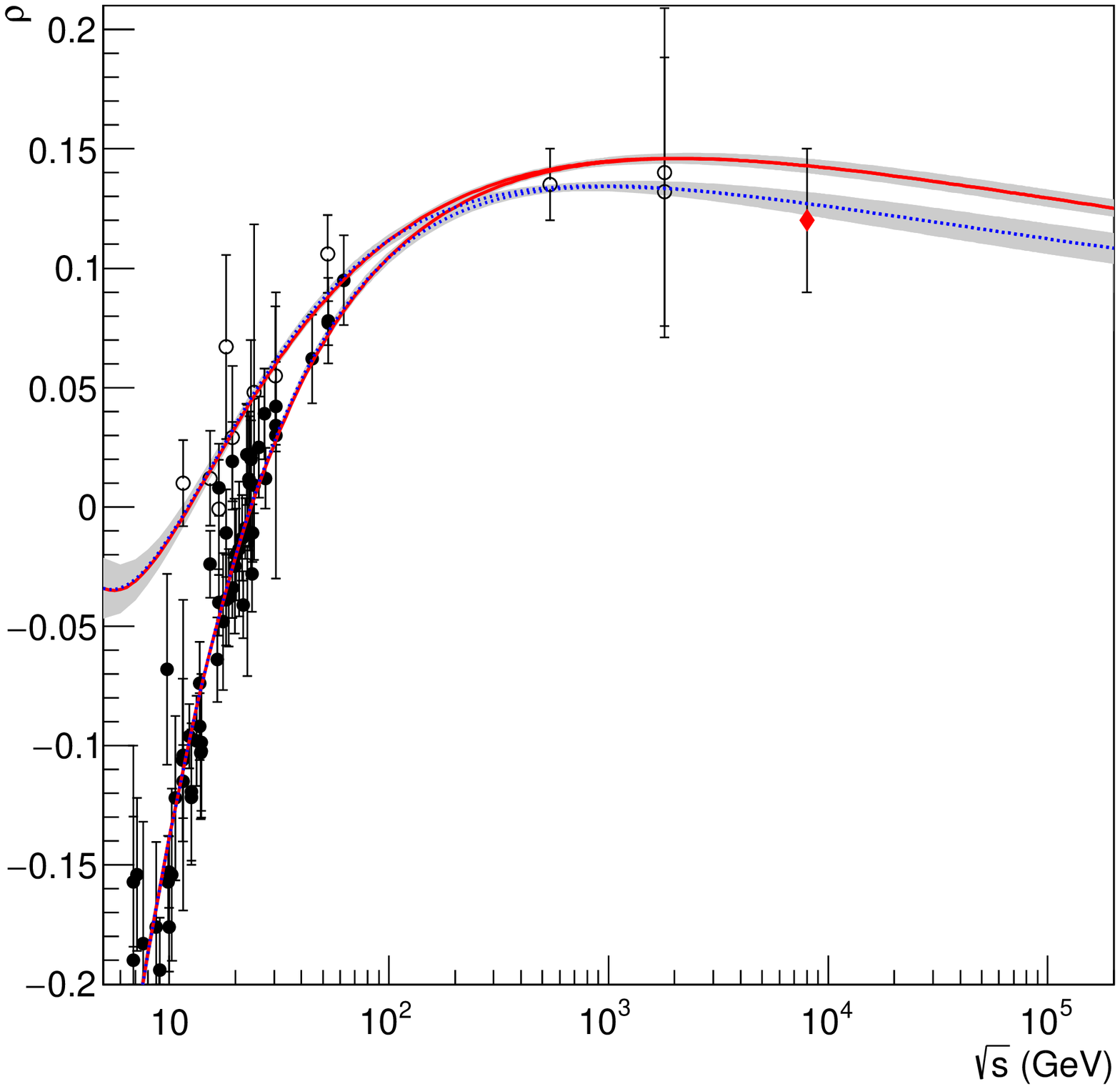,width=8cm,height=8cm}
\epsfig{file=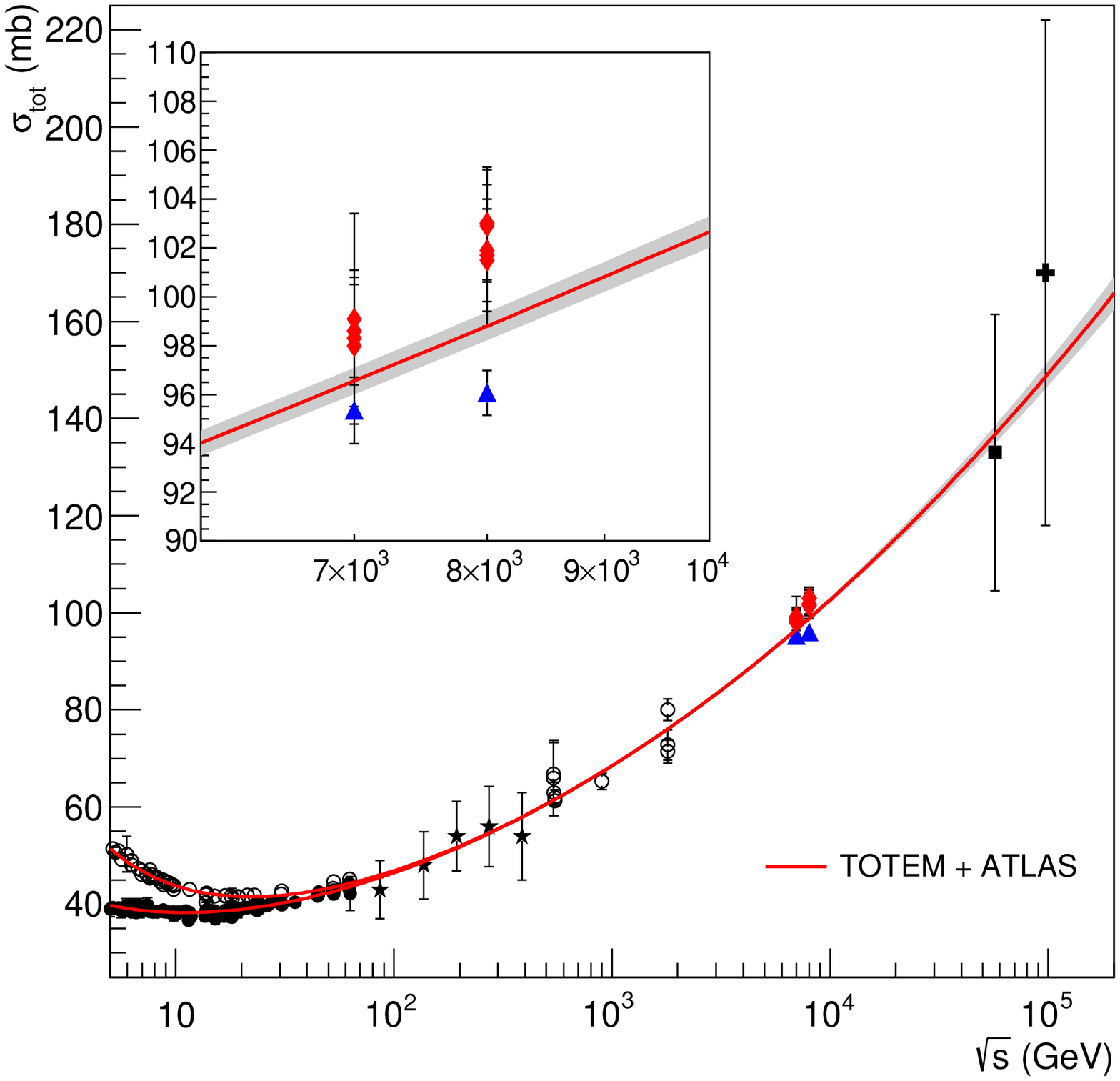,width=8cm,height=8cm}
\epsfig{file=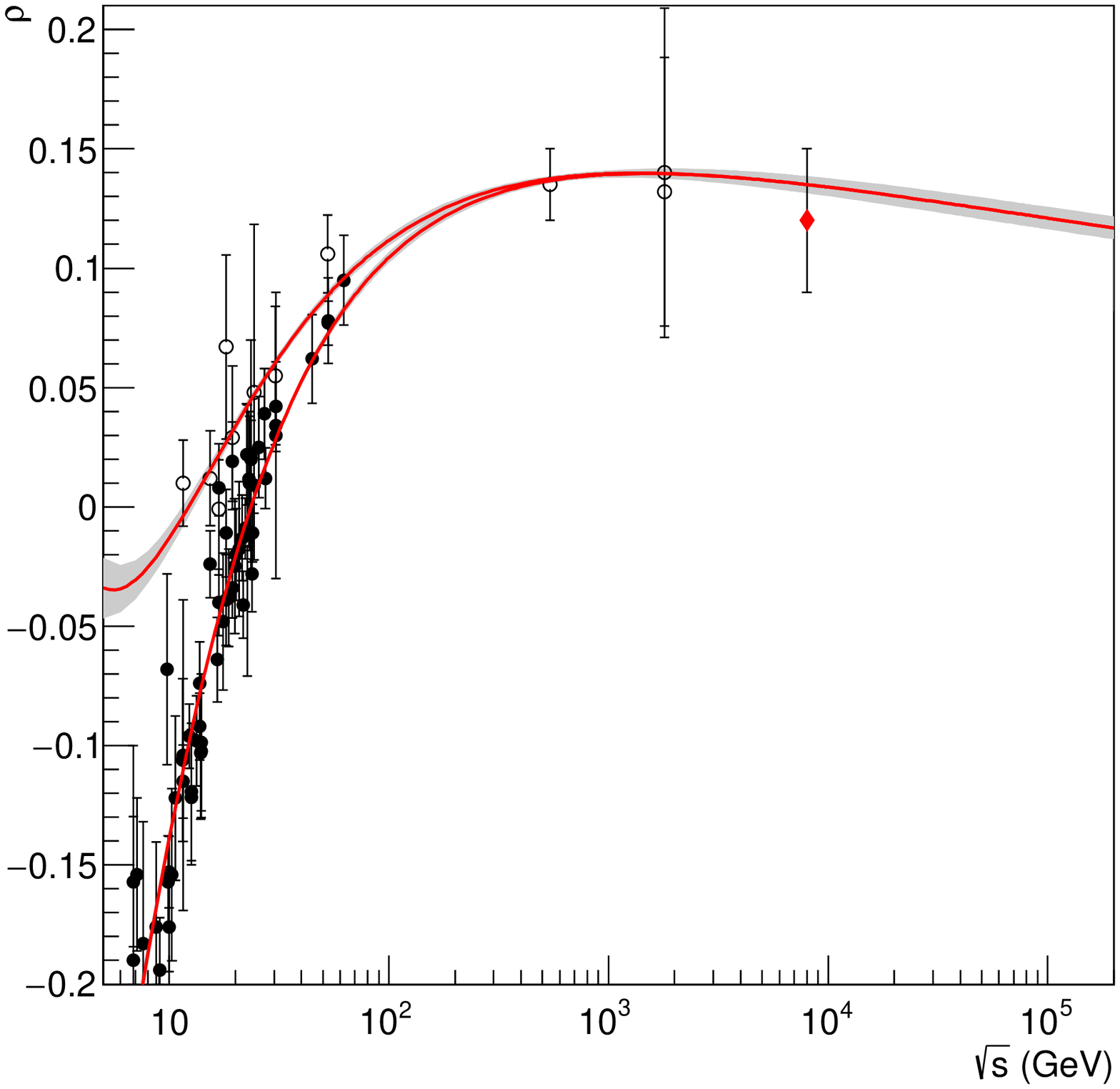,width=8cm,height=8cm}
\caption{Data reductions to $\sigma_{tot}$ (left) and $\rho$ (right) from $pp$ (black circles)
and $\bar{p}p$ (white circles) scattering. Fit results with the \textbf{L$\gamma$ model}
to ensembles T and A (above) and T+A (below).}
\label{f2}
\end{figure}

The fit results and statistical information on the goodness-of-fit with models L2 and L$\gamma$ to ensembles 
T, A and T+A are displayed in Table \ref{t2}.
For each ensemble, the corresponding curves (with the uncertainty regions from error propagation) and the experimental data
are shown in Figure \ref{f1} from model L2  and in Figure \ref{f2} from model L$\gamma$.

For each fit result (models L2, L$\gamma$ and ensembles T, A, T+A), the predictions for $\sigma_{tot}$
and $\rho$ at some energies of interest in $pp$ scattering are displayed in Tables \ref{t3} (L2 model)
and \ref{t4} (L$\gamma$ model). 

In this analysis all the TOTEM data have been considered as independent in the
data reductions, including the two points from \cite{totem5} and the two points
from \cite{totem6} (Table 1). See Appendix A for a discussion on this respect.

\begin{table}[H]
\centering
\caption{Fit results with models L2 and L$\gamma$ to ensembles T, A and T+A.
Parameters $a_1, a_2, \alpha, \beta$ in mb, $K_{eff}$ in mbGeV$^2$ and
$b_1, b_2, \gamma$ dimensionless.
Energy scale fixed, $s_0 = 4m_p^2 = 3.521$ GeV$^2$.} 
\vspace{0.1cm}
\begin{tabular}{c c c c c c c}
\hline
Ensemble: &\multicolumn{2}{c}{TOTEM}&\multicolumn{2}{c}{ATLAS} & \multicolumn{2}{c}{TOTEM + ATLAS}\\\cmidrule(lr){2-3} \cmidrule(lr){4-5} \cmidrule(lr){6-7} 
Model:          &    L2      & L$\gamma$ & L2         & L$\gamma$  & L2         & L$\gamma$          \\\hline
$a_1$           & 32.11(60)  & 31.5(1.3) & 32.39(86)  & 32.4(1.0)  & 32.16(67)  & 31.60(98)          \\
$b_1$           & 0.381(17)  & 0.528(57) & 0.435(19)  & 0.438(57)  & 0.406(16)  & 0.484(84)          \\
$a_2$           & 16.98(72)  & 17.10(74) & 17.04(72)  & 17.04(72)  & 17.01(72)  & 17.07(73)          \\
$b_2$           & 0.545(13)  & 0.546(13) & 0.545(13)  & 0.545(13)  & 0.545(13)  & 0.546(13)          \\
$\alpha$        & 29.25(44)  & 34.0(1.1) & 30.88(35)  & 31.0(2.1)  & 30.06(34)  & 32.8(2.2)          \\
$\beta$         & 0.2546(39) & 0.103(29) & 0.2347(35) & 0.231(83)  & 0.2451(28) & 0.151(71)          \\
$\gamma$        & 2 (fixed)  & 2.301(98) & 2 (fixed)  & 2.01(12)   & 2 (fixed)  & 2.16(16)           \\
$K_{eff}$       & 50(17)     & 109(36)   & 74(20)     & 75(27)     & 61(17)     & 90(42)             \\\hline 
$\nu$           & 242        & 241       & 235        & 234        & 244        & 243                \\
$\chi^2/\nu$    & 1.09       & 1.07      & 1.08       & 1.09       & 1.15       & 1.14               \\
$P(\chi^2)$     & 0.150      & 0.213     & 0.177      & 0.166      & 0.059      & 0.063              \\
\hline 
\end{tabular}
\label{t2}
\end{table}

%
%
%

\begin{table}[t]
\centering
\caption{Predictions for \tcs\ and \ro\ with the L2 model.} 
\vspace{0.1cm}
\begin{tabular}{c c c c c c c}
\hline
 &\multicolumn{2}{c}{TOTEM}&\multicolumn{2}{c}{ATLAS}&\multicolumn{2}{c}{TOTEM + ATLAS} \\\cmidrule(lr){2-3} \cmidrule(lr){4-5} \cmidrule(lr){6-7} 
 $\sqrt{s}$ (TeV) & \tcs (mb)   & $\rho$     & \tcs (mb)  & $\rho$      & \tcs (mb)  & $\rho$     \\
\hline
13               & 108.94(86)  & 0.1296(10) & 104.31(81) & 0.12489(92) & 106.76(61) & 0.12740(75)\\
14               & 110.28(88)  & 0.12915(96)& 105.55(83) & 0.12447(91) & 108.04(63) & 0.12694(74)\\
57               & 137.8(1.3)  & 0.11980(73)& 130.9(1.2) & 0.11625(69) & 134.51(92) & 0.11813(57)\\
95               & 148.8(1.5)  & 0.11646(66)& 141.0(1.4) & 0.11324(63) & 145.1(1.0) & 0.11494(52)\\
\hline
\end{tabular}
\label{t3}
\end{table}
\begin{table}[t]
\centering
\caption{Predictions for \tcs\ and \ro\ with the L$\gamma$ model.} 
\vspace{0.1cm}
\begin{tabular}{c c c c c c c}
\hline
 &\multicolumn{2}{c}{TOTEM}&\multicolumn{2}{c}{ATLAS}&\multicolumn{2}{c}{TOTEM + ATLAS} \\\cmidrule(lr){2-3} \cmidrule(lr){4-5} \cmidrule(lr){6-7} 
 $\sqrt{s}$ (TeV) & \tcs (mb)   & $\rho$     & \tcs (mb)  & $\rho$      & \tcs (mb)  & $\rho$     \\
\hline
13                & 110.7(1.2)  & 0.1417(47) & 104.34(97) & 0.1251(44) & 107.39(94) & 0.1333(63)\\
14                & 112.1(1.3)  & 0.1413(47) & 105.6(1.0) & 0.1247(44) & 108.8(1.0) & 0.1329(64)\\
57                & 143.4(2.8)  & 0.1337(52) & 131.0(2.2) & 0.1165(51) & 137.0(2.8) & 0.1251(73)\\
95                & 156.2(3.6)  & 0.1306(52) & 141.2(2.8) & 0.1135(52) & 148.4(3.7) & 0.1221(75)\\
\hline
\end{tabular}
\label{t4}
\end{table}


\section{Summary, Conclusions and Final Remarks}
\label{s4}

Attempting to quantify and investigate the consequences of discrepant measurements
of the total cross section by the TOTEM and ATLAS Collaborations, we have developed
6 fits to $\sigma_{tot}$ and $\rho$ data from $pp$ and $\bar{p}p$ scattering in
the energy region 5 GeV - 8 TeV. We applied the L2 and L$\gamma$ models (Section 2)
to three ensembles: T, A and T+A (Section 3). The fit results are presented in
Table \ref{t2} and Figures \ref{f1} and \ref{f2}. In what follows we first refer to the
results obtained with ensembles A and T and after that to those obtained with ensemble T+A.

Ensembles T and A indicate different scenarios for the rise of $\sigma_{tot}(s)$,
as shown in Figures \ref{f1} and \ref{f2}. These different behaviors can be quantified by
two parameters associated with model L$\gamma$ (Table \ref{t2}):
$\beta \sim$ 0.10 $\pm$ 0.03 mb, $\gamma \sim$ 2.3 $\pm$ 0.1 (T) and
$\beta \sim$ 0.23 $\pm$ 0.08 mb, $\gamma \sim$ 2.0 $\pm$ 0.1 (A).
From these results we can infer as extrema bounds for $\gamma$ the values
1.9 and 2.4.
However, with ensembles T or A and models L2 or L$\gamma$, the fits
provide $\chi^2/\nu \sim$ 1.1 and therefore the results can not be
distinguished on statistical grounds.

Let us focus now on the experimental data presently available, namely
ensemble T+A. 
The fit results with models L2 and L$\gamma$ are equivalent on statistical grounds:
$\chi^2/\nu = $ 1.15 (L2) and 1.14 (L$\gamma$), or $P(\chi^2) \sim $ 0.06
in both cases. From Figures 1 and 2 (L2 and L$\gamma$), the curves within the uncertainty
regions lie essentially between the TOTEM and ATLAS data. At 8 TeV, they reach the
lower end of the TOTEM uncertainties and lie above the ATLAS datum, namely this
point is not described by both models. For example, the results within this ensemble for $\sigma_{tot}$ at 8 TeV
are 98.58 $\pm$ 0.52 mb (L2) and 98.83 $\pm$ 0.60 mb (L$\gamma$) and
the differences between these predictions and the ATLAS datum, 
\begin{eqnarray} 
\frac{\sigma_{tot}^{\mathrm{Predicted}} - \sigma_{tot}^{\mathrm{ATLAS}}}{\Delta \sigma_{tot}^{\mathrm{ATLAS}}}, 
\nonumber
\end{eqnarray} 
read 2.7 (L2) and 3.0 (L$\gamma$).
With model L2 we obtained $\beta =$ 0.2451 $\pm$ 0.028 mb,
which is below the latest PDG value, $\beta_{\mathrm{PDG}} =$ 0.2720 $\pm$ 0.048 mb
\cite{pdg16}. With model L$\gamma$, $\beta =$ 0.151 $\pm$ 0.071 mb and
$\gamma =$ 2.16 $\pm$ 0.16. It is interesting to note that this $\gamma$-value
is in agreement with the result obtained by Amaldi et al., forty
years ago (data up to 62.5 GeV), namely $\gamma_{\mathrm{A}} =$ 2.10 $\pm$ 0.10 \cite{amaldi}.  

\begin{figure}[ht]
\centering
\epsfig{file=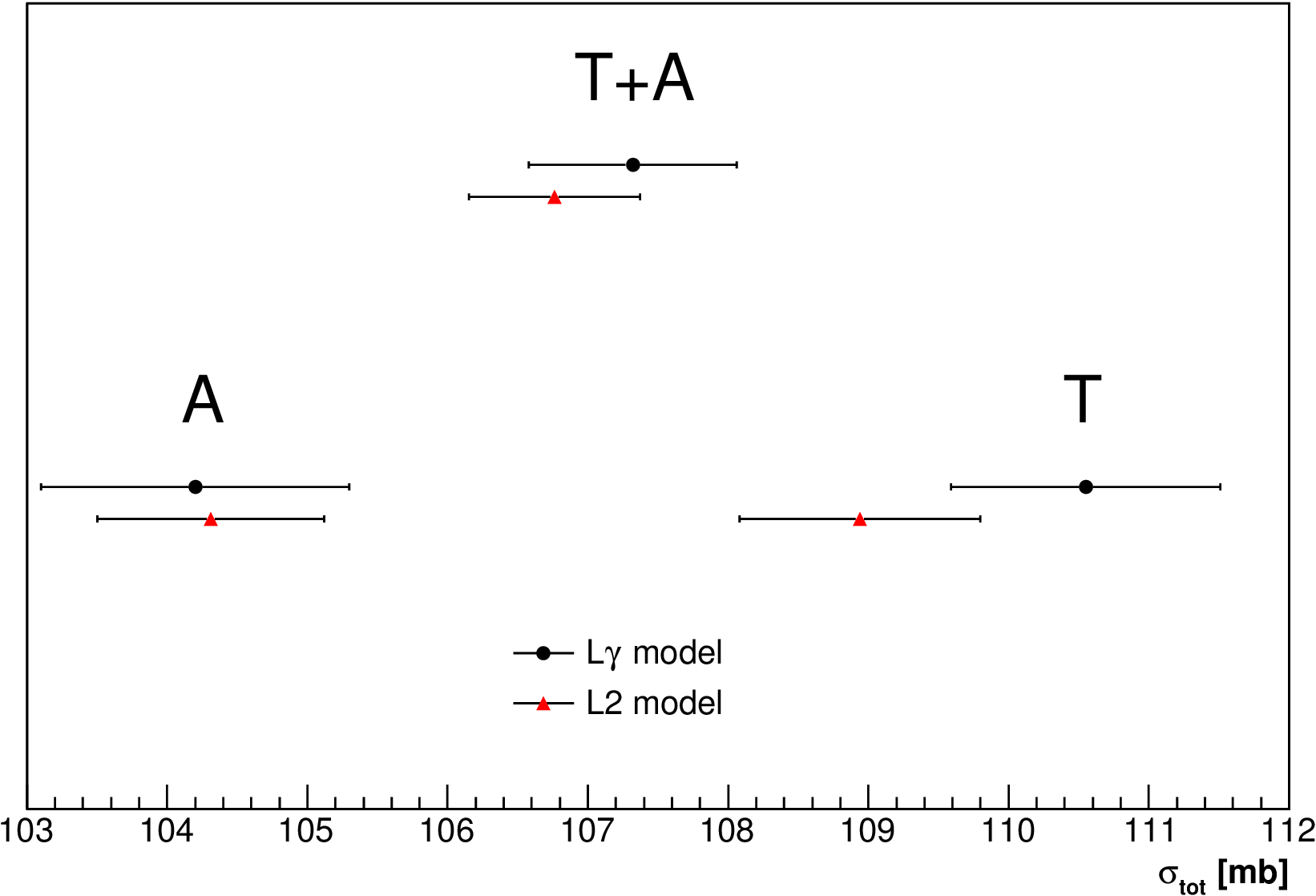,width=12cm,height=8cm}
\caption{Predictions for $\sigma_{tot}$ at 13 TeV within ensembles A, T+A and T
and models L2 and L$\gamma$.}
\label{f3}
\end{figure}

Within each one of the three ensembles, all the predictions for $\sigma_{tot}$ at 13 TeV with models L2 and L$\gamma$ (Tables
\ref{t3} and \ref{t4}) are in agreement within the uncertainties and cannot be distinguished. 
However, the results from distinct ensembles present differences, as illustrated in Figure
\ref{f3}.

In our L$\gamma$ model (with L2 as a particular case for $\gamma$ = 2), the
\textit{analytic} connection between $\sigma_{tot}(s)$ and $\rho(s)$ is obtained through singly
subtracted derivative dispersion relations, with the \textit{effective subtraction constant}
as a free fit parameter. Another analytic way to connect these two quantities is by
means of asymptotic uniqueness (Phragm\'en-Lindel\"off theorems) \cite{eden}, Section 7.1.
A detailed analysis, confronting these two analytic methods and the corresponding results
for $\sigma_{tot}(s)$ and $\rho(s)$ is presented in \cite{fms17b}.

Beyond providing useful parameters to quantify the different scenarios, the $\beta \ln^{\gamma} s$
leading term has also an interesting role in what concerns the rate of change of the total cross
section. As a function of the variable $\ln s$, the Froissart, Lukaszuk, Martin bound for
$\sigma_{tot}(\ln s)$ implies also
in two bounds related to rates of change, namely the slope ($S$) and the curvature ($C$) for $s \rightarrow \infty$: 
linear ($2\beta \ln s$) and constant ($2\beta$) limits, respectively.
On the other, the leading term predicts:
\begin{eqnarray}
S = \beta \gamma \ln^{\gamma - 1}s,
\qquad
\mathrm{and}
\qquad
C = \beta \gamma (\gamma - 1) \ln^{\gamma -2} s,
\nonumber
\end{eqnarray}
thus allowing detailed investigations on the rate of change of the
total cross section at high energies (not only $s \rightarrow \infty$).
For example, at 8 TeV and with the L$\gamma$ model, the fits with ensembles
T and A predict: $S \sim 9.3 \pm 0.2$ mb (T) and $S \sim 7.9 \pm 0.3$ mb (A), indicating a faster
rise of $\sigma_{tot}(s)$ from the TOTEM data than from the ATLAS data.

It is expected that further experimental analyses and the new data from LHC13
might clarify the different points and aspects raised in this work.

\vspace{0.3cm}

\section*{Acknowledgments}

We are thankful to David D. Chinellato for useful discussions. 
Research supported by FAPESP, Contract 2013/27060-3 (P.V.R.G.S.).

\appendix

\section{Comments on the Dataset and Fit Procedure}
\label{saa}

Among the TOTEM data at 8 TeV, two points published in \cite{totem5}
and two points in \cite{totem6} (see Table 1), correspond to
different analyses based on the same dataset. Specifically,
the two points from \cite{totem5} are related to quadratic or cubic polynomials
as parameterizations at the diffraction peak
and those from \cite{totem6} with central or peripheral phase formulations
(Coulomb-nuclear interference region).

Therefore, in a certain sense, the two values in each pair of points might not be considered 
as independent data, as assumed in our analysis. In order to test the influence
of this assumption (independent points), we have developed two re-analysis with 
the procedures that follow.

Firstly, we have estimated a mean value for each pair of points.
Since the systematic uncertainties are the same
(2.1 mb in \cite{totem5} and 2.3 mb in \cite{totem6}), 
in each case we have evaluated the arithmetic mean and the deviation
from the mean, which was then added to the corresponding systematic
uncertainties, resulting in two independent points:
$101.7 \pm 2.3$ mb from \cite{totem5} and $103.0 \pm 2.4$ mb from \cite{totem6}.
That seems a reasonable result since the deviation from the mean
have a systematic character and adding with the real systematic
uncertainty led to a larger uncertainty than that obtained by quadrature.
By substituting these two independent estimations in place of the four points
in ensembles T and T + A,
the same fit procedure (explained in the text) led to the results presented in Table \ref{t5}.

\begin{table}[ht]
\centering
\caption{Data reductions with the two TOTEM points from [27] and two from [28] (Table 1), 
substituted by their respective average estimations (see text). 
Fit results with models L2 and L$\gamma$ to ensembles T and T+A.
Parameters $a_1, a_2, \alpha, \beta$ in mb, $K_{eff}$ in mbGeV$^2$ and
$b_1, b_2, \gamma$ dimensionless.
Energy scale fixed, $s_0 = 4m_p^2 = 3.521$ GeV$^2$.} 
\vspace{0.1cm}
\begin{tabular}{c c c c c}
\hline
Ensemble: &\multicolumn{2}{c}{TOTEM}    & \multicolumn{2}{c}{TOTEM + ATLAS}\\\cmidrule(lr){2-3} \cmidrule(lr){4-5} 
Model:          & L2        & L$\gamma$ & L2        & L$\gamma$         \\\hline
$a_1$           & 32.11(61) & 31.5(1.3) & 32.21(69) & 31.72(90)         \\
$b_1$           & 0.389(18) & 0.522(57) & 0.414(16) & 0.472(92)         \\
$a_2$           & 16.99(71) & 17.09(72) & 17.02(72) & 17.06(73)         \\
$b_2$           & 0.545(13) & 0.546(13) & 0.545(13) & 0.546(13)         \\
$\alpha$        & 29.51(47) & 33.8(1.2) & 30.28(33) & 32.3(2.7)         \\
$\beta$         & 0.2516(45)& 0.109(32) & 0.2423(29)& 0.168(91)         \\
$\gamma$        & 2 (fixed) & 2.28(10)  & 2 (fixed) & 2.12(18)          \\
$K_{eff}$       & 53(18)    & 106(36)   & 64(18)    & 86(43)            \\\hline 
$\nu$           & 240       & 239       & 242       & 241               \\
$\chi^2/\nu$    & 1.09      & 1.08      & 1.12      & 1.12              \\
$P(\chi^2)$     & 0.151     & 0.196     & 0.089     & 0.089             \\
\hline 
\end{tabular}
\label{t5}
\end{table}

In a second procedure, we have used the four points, but attributing a statistical
weight equal to 1/2 to each one. The corresponding results are displayed in Table \ref{t6}.

\begin{table}[ht]
\centering
\caption{Data reductions attributing statistical weights equal to 1/2
to the four TOTEM points from [27] and [28] (Table 1). 
Fit results with models L2 and L$\gamma$ to ensembles T and T+A.
Parameters $a_1, a_2, \alpha, \beta$ in mb, $K_{eff}$ in mbGeV$^2$ and
$b_1, b_2, \gamma$ dimensionless.
Energy scale fixed, $s_0 = 4m_p^2 = 3.521$ GeV$^2$.} 
\vspace{0.1cm}
\begin{tabular}{c c c c c}
\hline
Ensemble: &\multicolumn{2}{c}{TOTEM}    & \multicolumn{2}{c}{TOTEM + ATLAS}\\\cmidrule(lr){2-3} \cmidrule(lr){4-5} 
Model:          & L2        & L$\gamma$ & L2         & L$\gamma$         \\\hline
$a_1$           & 32.11(63) & 31.5(1.3) & 32.20(69)  & 31.70(64)         \\
$b_1$           & 0.388(18) & 0.522(69) & 0.413(16)  & 0.473(57)         \\
$a_2$           & 16.99(72) & 17.09(73) & 17.02(72)  & 17.06(72)         \\
$b_2$           & 0.545(13) & 0.546(13) & 0.545(13)  & 0.546(13)         \\
$\alpha$        & 29.48(47) & 33.9(1.4) & 30.26(33)  & 32.4(1.6)         \\
$\beta$         & 0.2520(44)& 0.109(40) & 0.2427(29) & 0.166(55)         \\
$\gamma$        & 2 (fixed) & 2.28(13)  & 2 (fixed)  & 2.13(11)          \\
$K_{eff}$       & 53(18)    & 106(41)   & 64(18)     & 86(29)            \\\hline 
$\nu$           & 242       & 241       & 244        & 243               \\
$\chi^2/\nu$    & 1.09      & 1.07      & 1.12       & 1.12              \\
$P(\chi^2)$     & 0.170     & 0.221     & 0.099      & 0.099             \\
\hline 
\end{tabular}
\label{t6}
\end{table}

Taking into account the uncertainties in free fit parameters and the number
of degrees of freedom, comparison of Tables \ref{t5} and \ref{t6} shows that the two procedures led to
practically the same result.
Moreover, comparison with the results displayed in Table \ref{t2} (four points treated as
independents), shows that
all values of the free parameters are in agreement within the uncertainties and
the small differences in the goodness of fit are not relevant on statistical grounds. 

We conclude that, in the present case, the three procedures
led to numerical and statistical  results that are equivalent.

\end{document}